\documentclass[aps,twocolumn,amssymb,prl,showpacs,10pt]{revtex4-2}

\usepackage[utf8]{inputenc}
\usepackage{tabularx}
\usepackage{multirow}
\usepackage{url}
\usepackage{amsmath}
\usepackage{amssymb,amsmath}	
\usepackage{gensymb}
\usepackage{graphicx}
\usepackage{mathrsfs}
\usepackage{color}
\usepackage[normalem]{ulem}
\usepackage{float}
\usepackage{lmodern}
\usepackage{soul}
\usepackage[breaklinks,colorlinks,urlcolor=blue,citecolor=blue]{hyperref}

\usepackage{graphicx,verbatim}
\usepackage{stackengine}
\usepackage[caption=false]{subfig}
\usepackage{threeparttable}

\newcommand{\be}{\begin{equation}}
\newcommand{\ee}{\end{equation}}
\newcommand{\bi}{\begin{itemize}}
\newcommand{\ei}{\end{itemize}}
\newcommand{\figg}[1]{Fig.~\ref{fig:#1}}

\newcommand{\eq}[1]{Eq.~\ref{eq:#1}}

\newcommand{\fix}[1]{\textcolor{red}{[fix] }}

\newcommand{\comp}{\,c/\omega_{\mathrm{p}}}

\newcommand{\etarec}{\eta_{\rm rec}}

\newcommand{\tacc}{t_{\rm acc}}
\newcommand{\tesc}{t_{\rm esc}}
\newcommand{\omc}{\omega_{\rm c}}
\newcommand{\omcm}{\omega_{\rm c}^{-1}}

\begin{document}

\title{The origin of power-law spectra in relativistic magnetic reconnection}

\author{Hao Zhang}
\email{zhan2966@purdue.edu}
\affiliation{Department of Physics, Purdue University, West Lafayette, IN, 47907, USA }
\author{Lorenzo Sironi}
\email{lsironi@astro.columbia.edu}
\affiliation{Department of Astronomy and Columbia Astrophysics Laboratory, Columbia University, New York, NY 10027, USA}
\author{Dimitrios Giannios}
\affiliation{Department of Physics, Purdue University, West Lafayette, IN, 47907, USA}
\author{Maria Petropoulou}
\affiliation{Department of Physics, National and Kapodistrian University of Athens, University Campus Zografos, GR 15783, Greece}
\affiliation{Institute of Accelerating Systems \& Applications,
University Campus Zografos, GR 15783, Athens, Greece}

\date{\today}

\begin{abstract}
Magnetic reconnection is often invoked as a source of high-energy particles, and in relativistic astrophysical systems it is regarded as a prime candidate for powering fast and bright flares. We present a novel analytical model---supported and benchmarked with large-scale three-dimensional particle-in-cell simulations---that elucidates the physics governing the generation of power-law energy spectra in relativistic reconnection. Particles with Lorentz factor $\gamma\gtrsim 3\sigma$ (here, $\sigma$ is the  magnetization) gain most of their energy in the inflow region, while meandering between the two sides of the reconnection layer. Their acceleration time is $t_{\rm acc}\sim \gamma \,\eta_{\rm rec}^{-1}\omc^{-1}\simeq 20\,\gamma\,\omc^{-1}$, where $\eta_{\rm rec}\simeq0.06$ is the inflow speed in units of the speed of light and $\omc=eB_0/mc$ is the gyrofrequency in the upstream magnetic field. They leave the region of active energization after $t_{\rm esc}$, when they get captured by one of the outflowing flux ropes of reconnected plasma. We directly measure $t_{\rm esc}$ in our simulations and find that $t_{\rm esc}\sim t_{\rm acc}$ for $\sigma\gtrsim {\rm few}$. This leads to a universal (i.e., $\sigma$-independent) power-law spectrum $dN_{\rm free}/d\gamma\propto \gamma^{-1}$ for the particles undergoing active acceleration, and $dN/d\gamma\propto \gamma^{-2}$ for the overall particle population. Our results help shedding light on the ubiquitous presence of power-law particle and photon spectra in astrophysical non-thermal sources. 
\end{abstract}

\maketitle

Magnetic reconnection in relativistic plasmas \citep{lyutikov_uzdensky_03,lyubarsky_05,giannios_09,giannios_13,comisso_14} is invoked as a mechanism for efficient particle acceleration. It is the likely engine behind fast and bright flares observed from astrophysical non-thermal sources \citep[e.g.][]{cerutti_13a,yuan_16,lyutikov_18,petropoulou_16,nalewajko_19,christie_19,mehlhaff_20,hosking_sironi_20}, and a promising candidate for generating the ultra-high-energy cosmic rays (UHECRs) detected at Earth \citep[e.g.][]{zhang_sironi_21}. The process of particle acceleration in reconnection may be divided into three stages: (\textit{i}) the injection phase, that allows non-relativistic particles to be promoted to relativistic energies $\sim\sigma m c^2\gg mc^2$ ($\sigma$ is the magnetization, i.e., the ratio of magnetic to plasma enthalpy density); (\textit{ii}) the (potential) formation of a power-law spectrum of energetic particles; (\textit{iii}) particle acceleration up to the maximum energy (or, ``cutoff'') of the  spectrum.

Recent studies of relativistic reconnection ---largely based on fully-kinetic particle-in-cell (PIC) simulations \citep[e.g.][]{zenitani_01,ss_14,guo_14,werner_16,guo_19,zhang_sironi_21,werner_21}---have deepened our understanding of (\textit{i}) and (\textit{iii}). As regard to (\textit{i}), \cite{sironi_22} demonstrated that most of the particles ending up with high energies 
must have passed through regions where the assumptions of ideal magnetohydrodynamics are broken (however, see \cite{french_22}).
For (\textit{iii}), the rate of acceleration---or equivalently, the maximum energy attainable before particles cool or are advected out of the system---is dramatically different between 2D \cite{petropoulou_18,hakobyan_21} and 3D \cite{zhang_sironi_21}. While in 2D particles are buried within plasmoids / flux ropes in the reconnected plasma \footnote{In 2D, particle energization is governed by magnetic moment conservation in the increasing field of compressing plasmoids \cite{petropoulou_18,hakobyan_21}.}, in 3D particles with Lorentz factor $\gamma\gtrsim 3\sigma$
gain most of their energy in the inflow region, while meandering between the two sides of the reconnection layer.
 This results in fast acceleration, with $\gamma \propto t$ \cite{zhang_sironi_21}---in contrast, $\gamma\propto \sqrt{t}$ in 2D \cite{petropoulou_18,hakobyan_21}. Regarding 
 power-law formation (\textit{ii}), existing models \cite{zenitani_01,guo_14,uzdensky_20} are  based on 2D simulations. Given the 
 difference in particle dynamics and acceleration rates 
between 2D and 3D \footnote{Earlier works including \cite{ss_14} did not appreciate this difference, because the 3D domain was not large enough to follow particle acceleration to $\gamma\gg3\sigma$.}, the physics of power-law formation in 3D is unlikely to be a mere generalization of existing 2D theories.

In this {\it Letter}, we present the first analytical model of power-law formation in relativistic reconnection that self-consistently accounts for the 3D dynamics of high-energy particles. We benchmark our model with large-scale 3D PIC
 simulations, and demonstrate that at $\gamma \gtrsim 3\sigma$, relativistic reconnection leads to a $\sigma$-independent power-law spectrum $dN/d\gamma\propto \gamma^{-2}$.

\begin{figure}
\centering
    \includegraphics[width=\columnwidth]{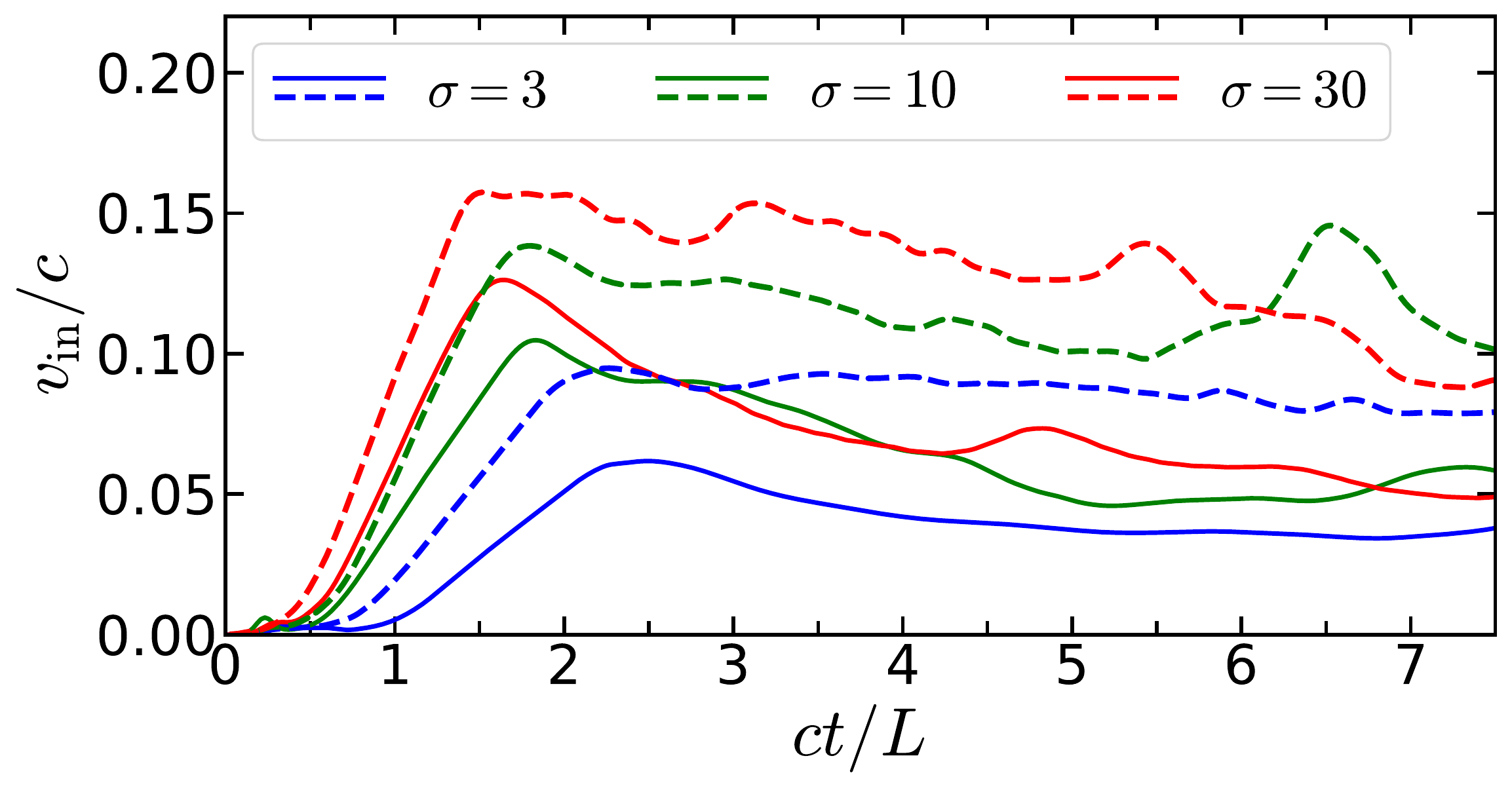}
    \caption{Time evolution of the reconnection rate $v_{\rm in}/c$, where $v_{\rm in}$ is the inflow speed averaged over $0.2L<y<0.6L$. Solid lines refer to 3D, dashed to 2D. The reconnection rate attains a quasi-steady value at $t\gtrsim 3\,L/c$.
    }
    \label{fig:recrate}
\end{figure}

\begin{figure*}
\centering   
    \includegraphics[width=0.75\textwidth]{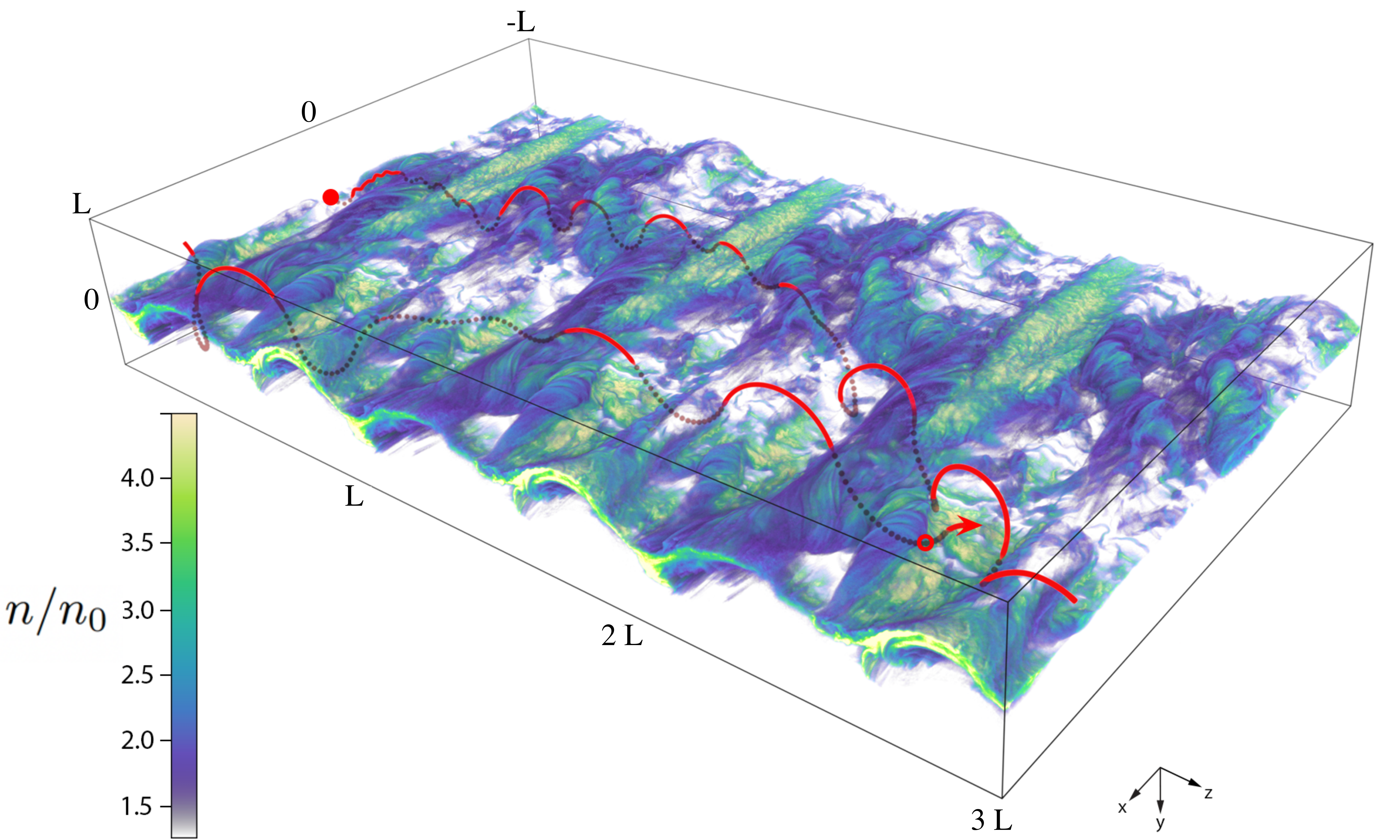}
    \caption{Isosurfaces of reconnected plasma density  from the 3D $\sigma=10$ simulation at $ct/L\simeq10$.
    The trajectory of a representative high-energy positron is overlaid, starting at the filled red circle and ending at the tip of the red arrow, after looping
once through the periodic $z$ boundary. It is colored in solid red if the positron is in the upstream (inflow) region above the midplane (on the same side as the observer), otherwise it is dotted black. The red open circle indicates the particle position at the time $ct/L\simeq10$ of the density isosurfaces. 
    The domain is replicated three times in $z$  for easier visualization of the positron track.}
    \label{fig:3dfld}
\end{figure*}

\begin{figure}
\centering   
    \includegraphics[width=\columnwidth]{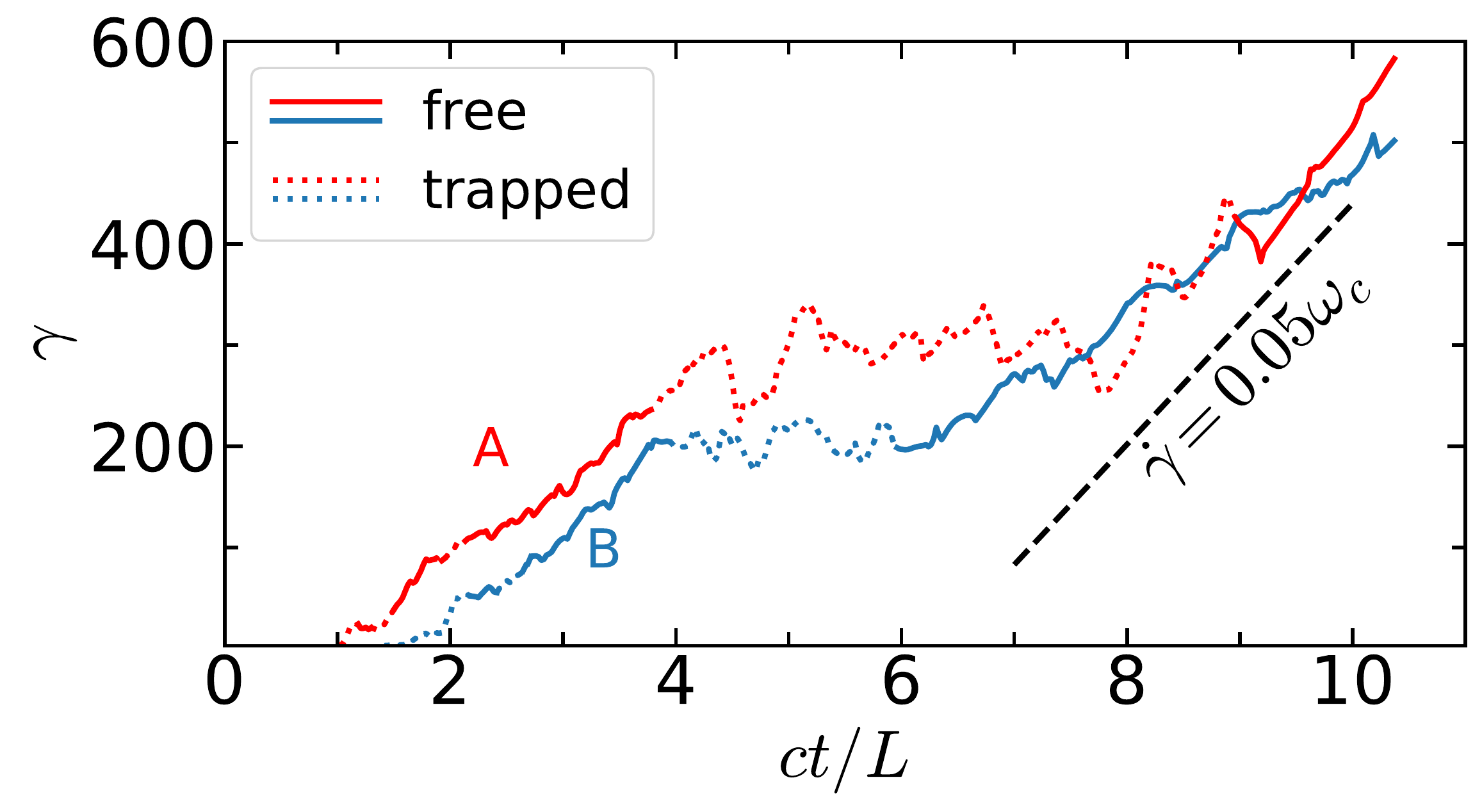}
    \caption{Time evolution of the Lorentz factor of two representative positrons from the 3D $\sigma=10$ simulation in \figg{3dfld}. Solid lines indicate when the positron is free, dotted when trapped, see text. The spatial track of positron A is in \figg{3dfld}.}
    \label{fig:3dprt}
\end{figure}

{\it Setup}---We build our model upon large-scale 3D PIC simulations performed with TRISTAN-MP \citep{buneman_93, spitkovsky_05}. We initialize a magnetic field of strength $B_0$ that reverses from $+\hat{x}$ to $-\hat{x}$ across a current sheet at $y=0$. We consider a cold electron-positron upstream plasma with {rest-frame} density $n_0$ of 2 particles per cell. 
The field strength $B_0$ is parameterized by the magnetization $\sigma = B_0^2 / 4\pi  n_0 m c^2 = \left(\omega_{\rm c} / \omega_{\rm p}\right)^2$, where $\omega_{\rm c} = e B_0 / m c$ is the gyrofrequency and $\omega_{\rm p} = \sqrt{4\pi n_0 e^2 / m}$ is the plasma frequency. We explore magnetizations $\sigma=3$, 10 and 30.
We also initialize a uniform ``guide'' field $B_g=0.1\,B_0$ along $z$.
Along the $y$-direction of inflows, two injectors continuously introduce fresh plasma and magnetic flux into the domain \citep{sironi_16}. We employ periodic boundary conditions in $z$ and outflow boundaries in $x$. We resolve the plasma skin depth $\comp$ with 2.5 cells for $\sigma=3$ and 10, and 2 cells for $\sigma=30$. We employ large domains, adopting $L=L_z=800\comp$ for $\sigma=3$ and 10, and $L=L_z=1400\comp$ for $\sigma=30$. Here, $L$ is the domain half-length along $x$, while $L_z$ is the domain length in $z$. Such large domains are essential to capture distinctive 3D effects \cite{zhang_sironi_21}. We also compare our results to 2D simulations having identical physical and numerical parameters (other than for $n_0$, which is 16 in 2D).

{\it Results}---We initiate reconnection by reducing the pressure of current-carrying particles near $(x,y)=(0,0)$ at the initial time \cite{sironi_16}. This generates two reconnection fronts, which propagate along $x$ and leave the domain at $t\sim 1.5 L/c$. The system settles into a statistical quasi-steady state at $t\gtrsim 3 L/c$, as indicated by 
the reconnection rate $v_{\rm in}/c$ (the inflow speed in units of the speed of light)  in \figg{recrate} (solid for 3D, dashed for 2D). The reconnection rate in 3D is lower than in 2D, and displays a weaker dependence on magnetization. In steady state, the 3D value averaged over $3.0\lesssim ct/L\lesssim 7.5$ is $\etarec\equiv v_{\rm in}/c\simeq 0.040$ for $\sigma=3$, $0.058$ for $\sigma = 10$, and $0.063$ for $\sigma = 30$.

A representative snapshot of  plasma density at late times is  in \figg{3dfld}. The reconnected plasma (hereafter, ``downstream'') is fragmented into plasmoids~/~flux ropes of
various sizes. 
The trajectory of a representative high-energy positron is overlaid.
At late times, the positron moves primarily along $z$ while performing  Speiser-like orbits 
\cite{speiser_65} that sample the inflow region (hereafter, ``upstream'') on both sides of the layer 
(\figg{3dfld}). The  energy history of the same positron is presented in \figg{3dprt} (particle A, red), together with another representative high-energy positron (particle B, blue). The energy tracks are marked with solid lines when the positrons are upstream (hereafter, ``free'' phase) and with dotted lines when downstream (hereafter, ``trapped'' phase, when particles are trapped within plasmoids). To distinguish between upstream and downstream, we define a ``mixing'' factor \cite{rowan_17,ball_18,sironi_beloborodov_20} $M \equiv 1-2|n_{\rm top}/n - 1/2|$, where $n_{\rm top}$ is the density of particles that started from $y>0$, while $n$ is the total density. We label $M<M_{\rm crit}=0.3$ as  upstream, and $M\geq0.3$ as  downstream. 
For each time $t$, we calculate the median of $M$ between $t - t_{\rm L}/2$ and $t + t_{\rm L}/2$, where $t_{\rm L}=2 \pi \gamma \, \omcm$ is the gyration time for a particle with Lorentz factor $\gamma$. If more than 50\% of median values from $t - t_{\rm L}/2$ to $t + t_{\rm L}/2$ are smaller than $M_{\rm crit}=0.3$, the particle is identified as free at time $t$, otherwise as trapped. 

\figg{3dprt} demonstrates that particles are rapidly accelerated during the free phase in the upstream (solid), whereas their energy stays nearly constant while trapped in the downstream (dotted). We find that most of the high-energy particles ($\sim 70\%$ if $\gamma\gtrsim 10\sigma$) experience at least one free phase in their life (Suppl.~Mat.), and 
most of their energy 
is acquired while in the upstream (\figg{3dprt}; see also Fig.~7(c) in \citep{zhang_sironi_21}).
In the upstream, the energization mechanism is drift acceleration via the grad-B speed from the field discontinuity across the layer \cite{giannios_10,lazarian_12}. The energy gain rate during the free phase approaches $\dot{\gamma}_{\rm acc}\simeq \eta_{\rm rec}\beta_z \omc\simeq 0.05\,\omc$ (dashed black in \figg{3dprt}), where $\eta_{\rm rec}\simeq 0.06$ is the reconnection rate for $\sigma=10$ (\figg{recrate}) and $\beta_z\simeq 0.8$ is the typical velocity of high-energy particles (normalized to $c$) along the $z$-direction of the  reconnection electric field
(Suppl.~Mat. and \cite{zhang_sironi_21}). 

\begin{figure}
\centering
    \includegraphics[width=\columnwidth]{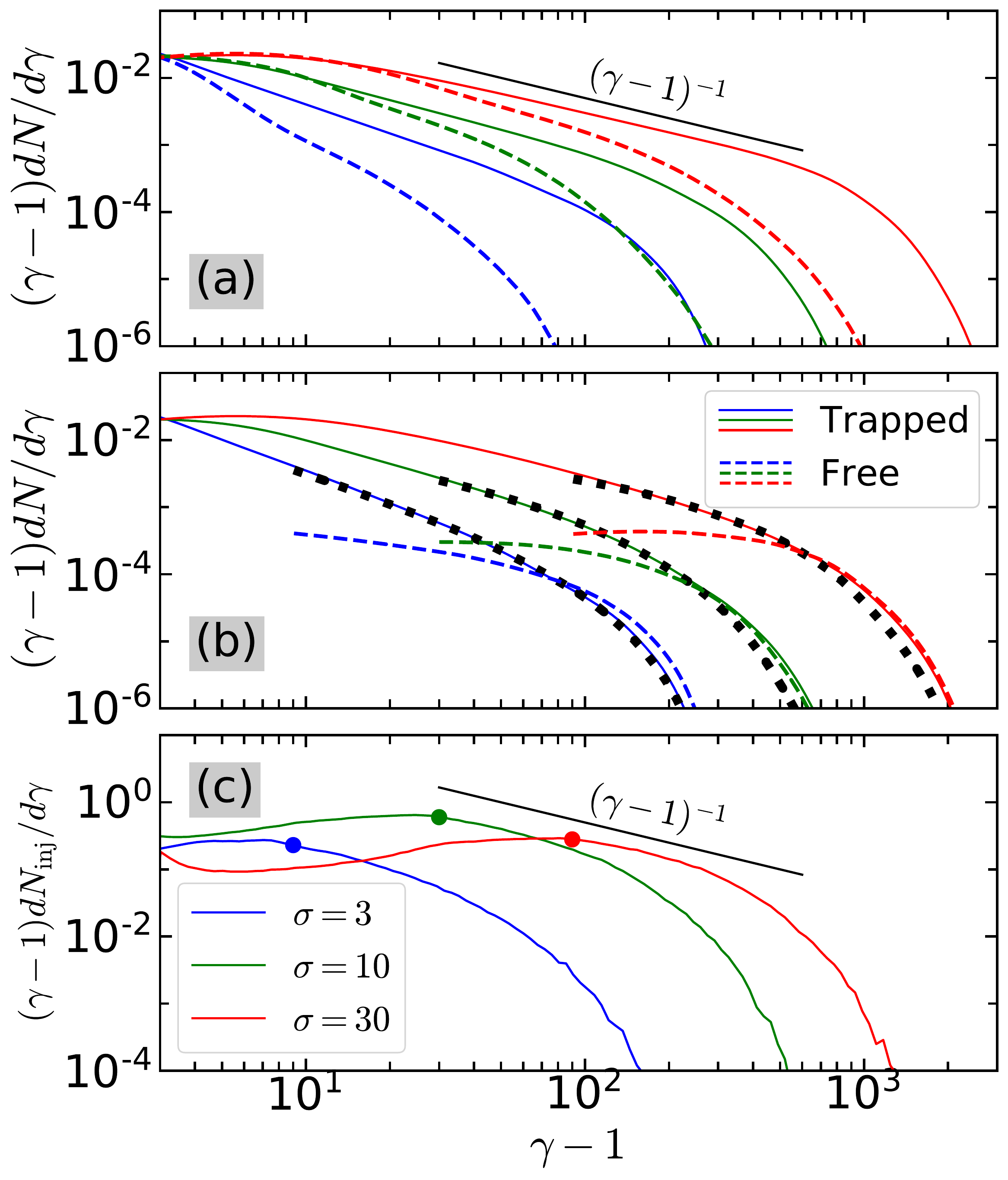}
    \caption{
    \textit{Top}: Particle energy spectra in 3D (solid) and 2D (dashed) for different magnetizations (legend in the bottom panel).
    \textit{Middle}: Spectra of free (dashed) and trapped (solid) particles. Free spectra are only shown for $\gamma\gtrsim 3\sigma$ (beyond injection). Black dotted lines show the predicted spectra of trapped particles based on \eq{predict}. In top and middle panels all curves intersect at $\gamma-1=3$. 
    \textit{Bottom}: Particle spectra measured at the injection time. We only consider particles whose maximum Lorentz factor is  $>3\sigma$, and we define injection as the time when a particle starts the free phase (each particle is counted more than once if it experiences more than one free phases).
 Spectra in the bottom panel are normalized such that their integral is unity.
     Spectra in all panels are time-averaged between $3\,L/c$ and $7.5\, L/c$. 
    }
    \label{fig:spect}
\end{figure}

\begin{figure}
\centering

    \includegraphics[width=\columnwidth]{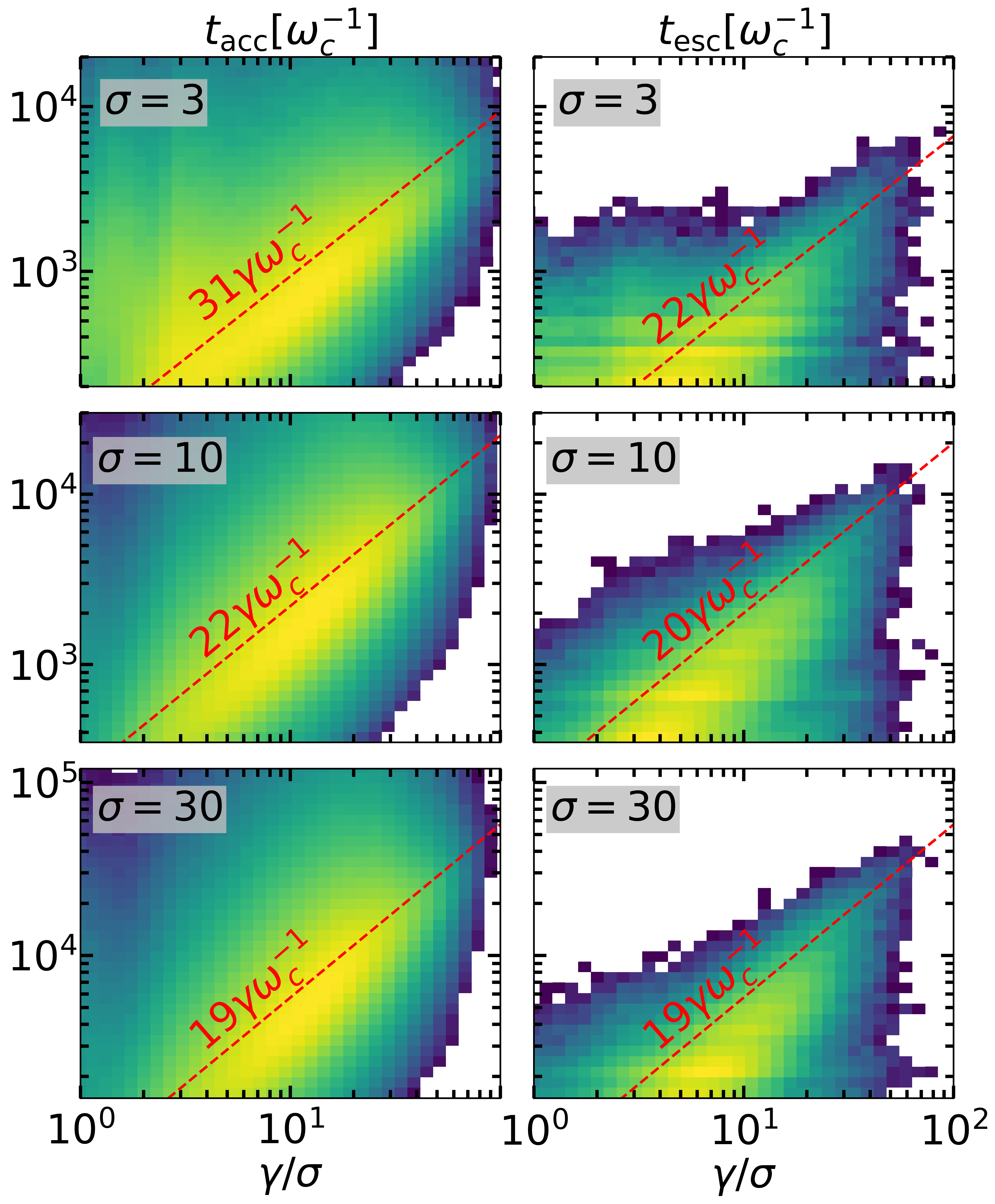}
    \caption{2D histograms of acceleration time $t_{\rm acc}$ (left) and escape time $t_{\rm esc}$ (right) of free particles. Dashed red lines are predictions based on the measured reconnection rate (left) and on the power-law slope of free particles (right), as described in the text. The Lorentz factor $\gamma$ on the horizontal axis is the instantaneous value on the left, while it is the value at the end of the free phase on the right. The histograms are normalized to their respective maxima, and colors span the range $[10^{-3},1]$ in logarithmic increments.
    }
    \label{fig:2dhist}
\end{figure}

The phase of free acceleration is artificially inhibited in 2D, where high-energy particles are buried in plasmoids
\citep{zhang_sironi_21}. This leads to a striking difference between 2D (dashed) and 3D (solid) spectra in \figg{spect}(a).  2D spectra are systematically steeper,  cut off at lower energies, and show a stronger dependence on $\sigma$. The 3D spectrum above $\gamma\sim 3\,\sigma$ can be modeled as a power law, whose slope $s\equiv-d\log N/d\log (\gamma-1)$ is nearly independent of the magnetization: $s\simeq 2.4$ for $\sigma=3$, $s\simeq 2.1$ for $\sigma=10$ and $s\simeq 2.0$ for $\sigma=30$. The power-law range extends up to a cutoff energy $\propto L$, such that the particle Larmor radius at the cutoff is comparable to the size of the largest plasmoids, $\sim 0.1 \,L$ \citep{zhang_sironi_21}.
The spectrum $d N_{\rm trap}/d\gamma$ of trapped particles  (solid in \figg{spect}(b)) is nearly identical to the overall spectrum $d N_{}/d\gamma$ (solid in \figg{spect}(a)), while the spectrum $d N_{\rm free}/d\gamma$ of free particles (dashed in \figg{spect}(b)) is harder. Near 
the high-energy cutoff, free and trapped particles contribute roughly equally.

We now present an analytical model for the power-law slope of the 3D particle spectrum. For each 3D simulation, we analyze the trajectories of $10^6$ positrons. Their spectrum at injection---at the beginning of the free phase---is shown in \figg{spect}(c). The highest energy particles may experience more than one free phase, as in \figg{3dprt}---after they get trapped, they break free again. {In this case,} the spectra in \figg{spect}(c) treat the beginning of each free stage as a separate injection episode. Most of the particles get injected at $\gamma_{}\sim 3\sigma$ (filled circles in \figg{spect}(c)). The injection spectrum at higher energies is steeper than $(\gamma-1)dN/d\gamma\propto (\gamma-1)^{-1}$ (solid black line), so our conclusions are the same as in the simple case of monoenergetic injection at $\gamma_{\rm inj}=3\sigma$  (see Suppl.~Mat.), which we adopt in the following. 

In steady state the distribution $f_{\rm free}=d N_{\rm free}/d\gamma$ of free particles 
{is governed by}
\begin{equation}\label{eq:fp}
    \frac{\partial}{\partial \gamma} \left(\dot{\gamma}_{\rm acc}f_{\rm free}\right) + \frac{f_{\rm free}}{t_{\rm esc}} = Q_{\rm inj} \delta(\gamma-\gamma_{\rm inj}),
\end{equation}
where injection into the free phase occurs at $\gamma_{\rm inj}=3\sigma$ with rate $Q_{\rm inj}$.
Following injection, we assume that particles experience fast acceleration with rate $\dot{\gamma}_{\rm acc}\simeq \eta_{\rm rec}\beta_z \omc$ during  the free phase in the upstream, {while} 
 no significant energization happens  in the downstream. The free phase terminates after $t_{\rm esc}$, when they get trapped and so leave the upstream region of active acceleration.
If both $\tacc\equiv\gamma/\dot{\gamma}_{\rm acc}$ and $\tesc$ scale linearly with $\gamma$ (as we demonstrate below), the solution is a power law \cite{kirk_98}
\begin{equation}
\label{eq:slope}
\frac{dN_{\rm free}}{d\gamma}\propto \gamma^{-s_{\rm free}} ~~~{\rm with}~~~ s_{\rm free}=t_{\rm acc}/t_{\rm esc}~.
\end{equation}

We measure $\tacc$ and $\tesc$  for $10^6$ particles in each of our 3D simulations, see \figg{2dhist}. 
Dashed red lines in the left column indicate the scaling expected for $\tacc$ if $\dot{\gamma}_{\rm acc}\simeq \beta_z \eta_{\rm rec} \omc$ with $\beta_z=0.8$, which provides a good fit to the locus of maxima (in yellow) of the 2D distributions. The escape time $\tesc$ (right column  in \figg{2dhist}) is the duration of each free phase (potentially more than one, for a given particle). According to \eq{slope}, we expect $\tesc\simeq \tacc/s_{\rm free}$, where $s_{\rm free}$ is the best-fit slope of the power-law range of the free  spectrum in \figg{spect}(b) (dashed), yielding $s_{\rm free}\simeq 1.4$ for $\sigma=3$, 1.1 for $\sigma=10$, and $1.0$ for $\sigma=30$. The expected $\tesc\simeq \tacc/s_{\rm free}$ is indicated with dashed  red lines in the right column of \figg{2dhist}, showing good agreement with the locus of maxima of the 2D histograms. Thus, our measurements of $\tacc$ and $\tesc$ in \figg{2dhist} are consistent with the slope $s_{\rm free}=\tacc/\tesc$
 of the free spectrum (dashed in \figg{spect}(b)).

We can finally relate the spectra of free  and trapped particles. In steady state,  the rate at which free particles get trapped should be equal---at each $\gamma$---to the rate at which trapped particles advect out of the $x$ boundaries. This yields
\begin{equation}
\frac{dN_{\rm trap}}{d\gamma}=\frac{t_{\rm adv}}{t_{\rm esc}} \frac{dN_{\rm free}}{d\gamma}~. 
\label{eq:predict}
\end{equation}
We measure the advection time $t_{\rm adv}$ from low-energy ($\gamma\lesssim 3\sigma$) particles that never experience a free stage and find $t_{\rm adv}\sim 1.5 \,L/c$. 
Using the free particle spectrum (dashed in \figg{spect}(b)) and \eq{predict}, we derive the dotted black lines in \figg{spect}(b), which overlap nearly perfectly with the  trapped particle spectrum (solid in \figg{spect}(b)) \footnote{More precisely, we use $t_{\rm adv}=1.3L/c$ for $\sigma=3$, $2.0L/c$ for $\sigma=10$ and $1.5L/c$ for $\sigma=30$.}. This implies that, even though at any given time the number of free particles  is much smaller (by a factor $\sim \tesc/t_{\rm adv}$) than the number of trapped particles (compare dashed and solid in \figg{spect}(b)), nearly all the $\gamma\gtrsim 3\sigma$ particles that are currently trapped had one (or more) prior episodes of fast
acceleration as free particles, during which they acquired
most of their energy.
Given that $\tesc\propto \gamma$ while $t_{\rm adv}$ is independent of $\gamma$, the slope of the free spectrum and of the trapped spectrum---which is the same as the one of the overall spectrum---are related by $s_{\rm free}=s-1$.
Since $s_{\rm free}\simeq 1$, then $s\simeq 2$, nearly independent of $\sigma$.

{\it Conclusions}---We present an analytical model---benchmarked with large-scale PIC simulations--- 
for power-law formation in relativistic reconnection
that self-consistently accounts for the 3D dynamics of high-energy
particles. 
High-energy particles gain most of their energy in a ``free'' phase spent in the upstream. In pair plasmas injection into the free phase occurs at $\gamma\sim3\sigma$, while in electron-proton plasmas at $\gamma_p\sim 3\sigma$ for protons and $\gamma_e\sim 3(m_p/m_e)\sigma$ for electrons \footnote{It is well known that $\sigma\gg1$ reconnection behaves similarly in electron-positron, electron-proton \citep{guo_16b,werner_18,ball_18} and electron-positron-proton \citep{petropoulou_19} plasmas.}.

The acceleration rate of free particles is energy-independent and approaches the maximum value $\simeq e E_{\rm rec} c$ associated with the reconnection electric field $E_{\rm rec}\simeq 0.06 B_0$. Fast  acceleration continues until the particles leave the layer or radiative cooling becomes important. Protons in powerful AGN jets can reach 
ultra-high 
energies $\sim10^{20}\rm eV$ \cite{giannios_10,zhang_sironi_21}.  

During the free phase, the acceleration time  $t_{\rm acc} \propto \gamma$ is comparable to the time $t_{\rm esc}\propto \gamma$ the particles spend in the free phase before getting trapped within plasmoids.  This yields  
a universal (nearly $\sigma$-independent) power-law spectrum $dN_{\rm free}/d\gamma\propto \gamma^{-1}$ for  the free particles, and $dN/d\gamma\propto \gamma^{-s}$ with $s\simeq 2$ for the overall particle population. Electron spectra with $s\gtrsim 2$ are commonly invoked in modeling the emission of AGN jets \cite[e.g.,][]{celotti_08,tavecchio_10}, while proton spectra with similar slopes may be required to explain simultaneously the spectrum and composition of UHECRs above $\sim 5\,\rm EeV$~\citep[e.g.,][]{2019FrASS...6...23B,2021EPJC...81...59D}.

Further work is needed to generalize our results to the regime of strong guide fields and to the case in which fast cooling losses lead to denser and smaller plasmoids, thus changing their cross-sectional area and reducing their probability of capturing free particles.

\phantom{xx}
\begin{acknowledgments}
We thank E. Nakar for insightful comments. L.S. acknowledges support from the Cottrell Scholars Award, NSF AST-2108201, and the DoE Early Career Award DE-SC0023015. This research was facilitated by the Multimessenger Plasma Physics Center (MPPC), NSF grant PHY-2206607. This project made use of the following computational resources: NASA Pleiades supercomputer, Habanero and Terremoto HPC clusters at Columbia University.
\end{acknowledgments}

\bibliographystyle{apsrev}
\bibliography{blob.bib}

\section{Supplemental Material}
\subsection{Dependence on the injection spectrum}
The distribution function
of free particles with Lorentz factor $\gamma$, i.e., $f_{\rm free}\equiv dN_{\rm free}/d\gamma$, evolves according to 
\begin{equation} \label{eq:diff}
    \frac{\partial f_{\rm free}}{\partial t} + \frac{\partial}{\partial\gamma}\left(\dot{\gamma}_{\rm acc}f_{\rm free} \right) + \frac{f_{\rm free}}{t_{\rm esc}} = Q_{\rm inj} \delta(\gamma-\gamma_{\rm inj}),
\end{equation}
where we assume monoenergetic injection at $\gamma_{\rm inj}\sim 3\sigma$. Here, $\dot{\gamma}_{\rm acc}=\eta_{\rm rec}\beta_{z}\omc$ is the acceleration rate and $t_{\rm esc}$ is the escape time from the acceleration region, i.e., the time that free particles spend in the upstream before getting trapped by plasmoids. As demonstrated in the main paper, both the acceleration and escape times scale linearly with the particle Lorentz factor, namely $\tacc\equiv\gamma/\dot{\gamma}_{\rm acc} \equiv t_{\rm acc,0}\gamma$ and 
$\tesc \equiv t_{\rm esc,0} \gamma$, where $t_{\rm acc,0}$ and $t_{\rm esc,0}$ do not depend on $\gamma$.
Assuming constant injection in time, the solution of \eq{diff} reads:
\begin{equation}\label{eq:solSimp}
f_{\rm free} =  Q_{\rm inj} t_{\rm acc,0} \left(\frac{\gamma}{\gamma_{\rm inj}} \right)^{-s_{\rm free}}, {\;\rm for\;} \gamma_{\rm inj} < \gamma < \gamma_{\rm inj} + t/t_{\rm acc,0}.
\end{equation}
The maximum energy of the distribution cannot grow indefinitely, but will stop at $\gamma=\gamma_{\rm cut}$ when  free particles cannot be confined anymore in the system (the so-called Hillas criterion). Therefore, \eq{solSimp} describes a power law of slope $s_{\rm free}=t_{\rm acc,0}/t_{\rm esc,0}$ for $\gamma_{\rm inj}<\gamma \ll \gamma_{\rm cut}$.

So far we have assumed monoenergetic injection. If particles are injected with a power-law distribution, we can replace the $\delta$-function in the right-hand side of Eq.~\ref{eq:diff} with a more general distribution function $f_{\rm inj}(\gamma)$
\begin{equation} \label{eq:powlawdiff}
    \frac{\partial f_{\rm free}}{\partial t} + \frac{\partial}{\partial\gamma}\left(\dot{\gamma}_{\rm acc}f_{\rm free} \right) + \frac{f_{\rm free}}{t_{\rm esc}} = f_{\rm inj},
\end{equation}
where
\begin{equation}
    f_{\rm inj} = \left
    \{\begin{array}{ll}
        0, & \gamma < \gamma_{\rm inj} \\
        Q_{\rm inj}\gamma^{-p}, & \gamma \geq \gamma_{\rm inj} \\
    \end{array}
    \right..
\end{equation}
The Green's function of Eq.~\ref{eq:powlawdiff} satisfies
\begin{equation} \label{eq:greens}
    \frac{\partial G}{\partial t} + \frac{\partial}{\partial\gamma}\left(\dot{\gamma}_{\rm acc}G \right) + \frac{G}{t_{\rm esc}} =  \delta(\gamma-\gamma_{\rm s}),
\end{equation}
which is  identical to Eq.~(\ref{eq:diff}) if the injection rate is set to unity. Therefore, the Green function (in steady state) for $\gamma_{\rm cut} \gg \gamma_{\rm s}$ reads
\begin{equation}
    G(\gamma, \gamma_s) \approx t_{\rm acc,0} \left(\frac{\gamma}{\gamma_{\rm s}}\right)^{-s_{\rm free}}.
\end{equation}
The solution to Eq.~(\ref{eq:powlawdiff}) can be then written as
\begin{equation}
\resizebox{\linewidth}{!}{$
    f_{\rm free} = \int ^{\infty} _{\gamma_{\rm inj}} G(\gamma, \gamma_s) f_{\rm inj}(\gamma_s) d\gamma_s = Q_{\rm inj}t_{\rm acc,0} \gamma^{-s_{\rm free}} \int ^{\gamma} _{\gamma_{\rm inj}} \gamma_s^{s_{\rm free}-p} d\gamma_s
$}
\end{equation}
Let us assume $\gamma \gg \gamma_{\rm inj}$. We find that if $p < s_{\rm free} + 1$, the solution becomes
\begin{equation}\label{eq:f_free_solution1}
    f_{\rm free} = \frac{Q_{\rm inj}t_{\rm acc,0}}{s_{\rm free}-p+1} \gamma^{-p+1} \propto \gamma^{-p+1}.
\end{equation}
Instead, if $p > s_{\rm free} + 1$, the solution is
\begin{equation}\label{eq:f_free_solution2}
    f_{\rm free} = \frac{Q_{\rm inj}\gamma_{\rm inj}^{s_{\rm free}-p+1}t_{\rm acc,0}}{-s_{\rm free}+p-1} \gamma^{-s_{\rm free}}
    \propto \gamma^{-s_{\rm free}}.
\end{equation}
The free particle spectrum is therefore the same as in the case of monoenergetic injection 
if $p > s_{\rm free} + 1$. Indeed, this condition is met
in our simulations, as shown by Fig.~4(c) in the main paper.

\begin{figure}
\centering
    \includegraphics[width=\columnwidth]{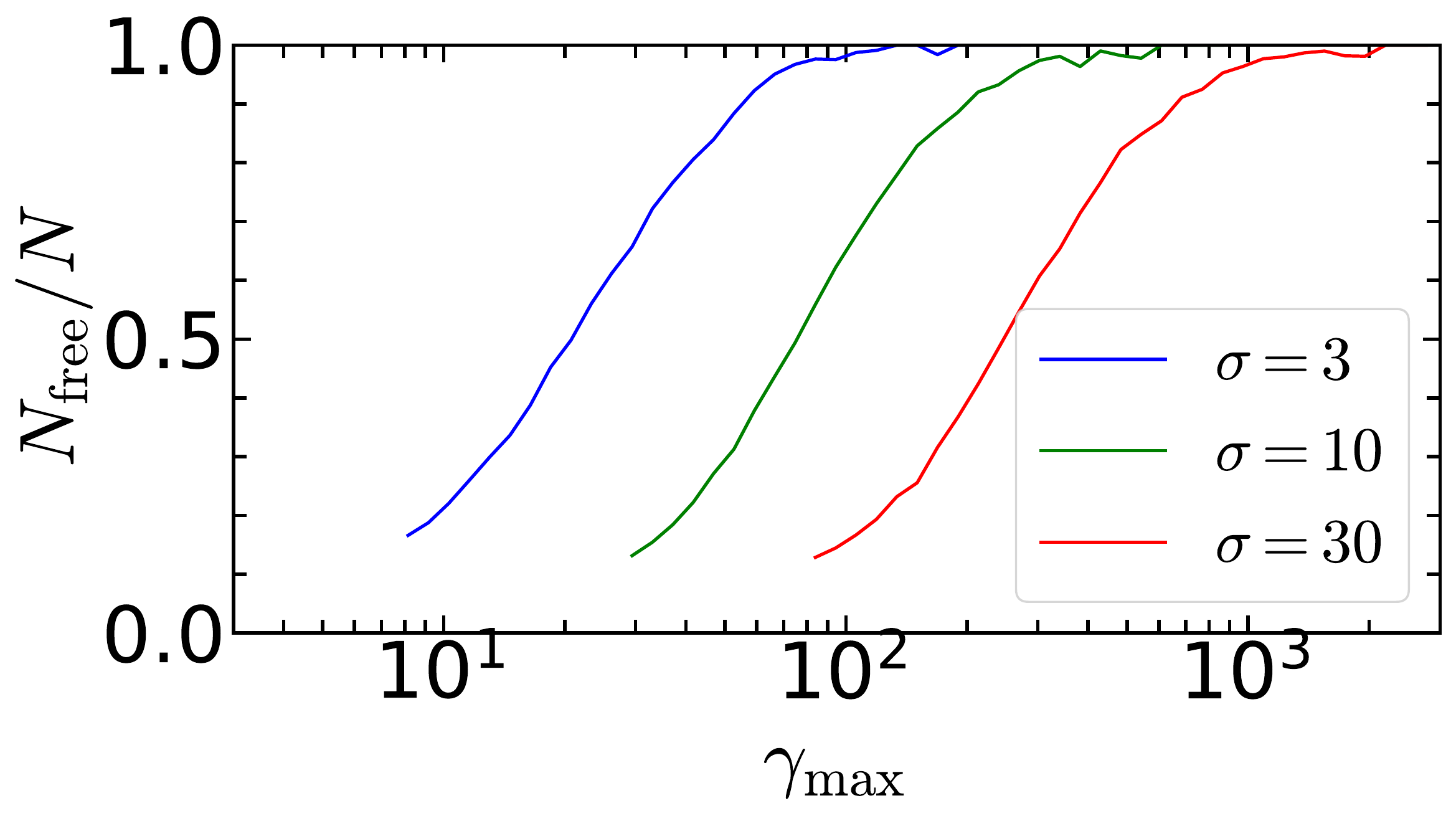}
    \caption{Fraction of particles that have experienced at least one free phase, as a function of the maximum Lorentz factor $\gamma_{\rm max}$ attained during the particle life. We only show the $\gamma\gtrsim 3\sigma$ range beyond injection.}
    \label{fig:frac}
\end{figure}

\begin{figure}
\centering
    \includegraphics[width=\columnwidth]{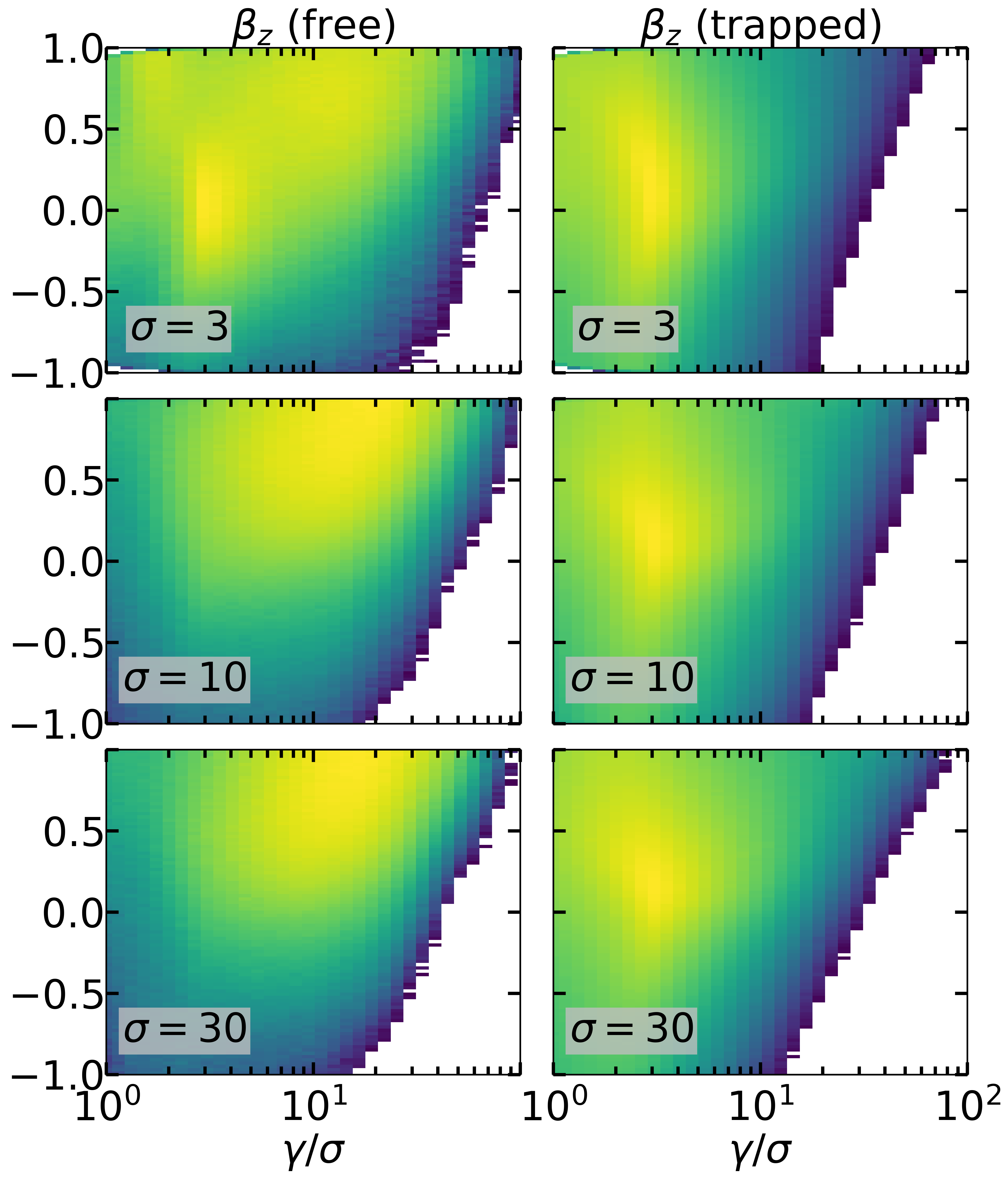}
    \caption{2D histograms of $\beta_z$, for free (left) and trapped (right) positrons (the corresponding figure for electrons is mirror-symmetric with respect to the $\beta_z=0$ axis).
    The Lorentz factor $\gamma$ on the horizontal axis is the instantaneous value. The histograms are normalized to their respective maxima, and colors span the range $[10^{-3},1]$ in logarithmic increments.
}
    \label{fig:betaz}
\end{figure}

\subsection{Fraction and $z$-velocity of free particles}
In \figg{frac} and \figg{betaz} we quantify some of the properties of free particles. \figg{frac} shows the fraction of particles that have experienced at least one free phase in their life, as a function of the particle maximum Lorentz factor $\gamma_{\rm max}$. We point out that this is different than the fraction of particles that are in the free phase at any given time. The latter is much smaller than the former, since the free phase is much shorter than the trapped phase (by a factor $t_{\rm esc}/t_{\rm adv}$, which is much smaller than unity for all particles well below the spectral cutoff). The figure shows that $\sim 20\%$ of particles with $\gamma_{\rm max}\sim 3\sigma$ have experienced at least one stage of free acceleration (this becomes $\sim 30\%$ when considering the cumulative contribution of all particles with $\gamma_{\rm max}\gtrsim 3\sigma$). This fraction increases to $\sim 70\%$ for particles with $\gamma_{\rm max}\sim 10\sigma$. These estimates are remarkably independent of $\sigma$---when using $\gamma_{\rm max}/\sigma$ on the horizontal axis, the three curves nearly overlap.

In \figg{betaz}, we present 2D histograms of the dimensionless $z$-velocity $\beta_z$, for free (left) and trapped (right) positrons. We show that trapped particles have $\beta_z\sim 0$, whereas free positrons tend to move nearly along the $+\hat{z}$ direction, i.e., with $\beta_z$ close to unity. Since the reconnection electric field also lies along $+\hat{z}$, the velocity of free positrons has optimal orientation for rapid acceleration. In the text, we take a typical value $\beta_z=0.8$ for our estimates of the acceleration rate of free particles.

\begin{figure}
\centering
    \includegraphics[width=\columnwidth]{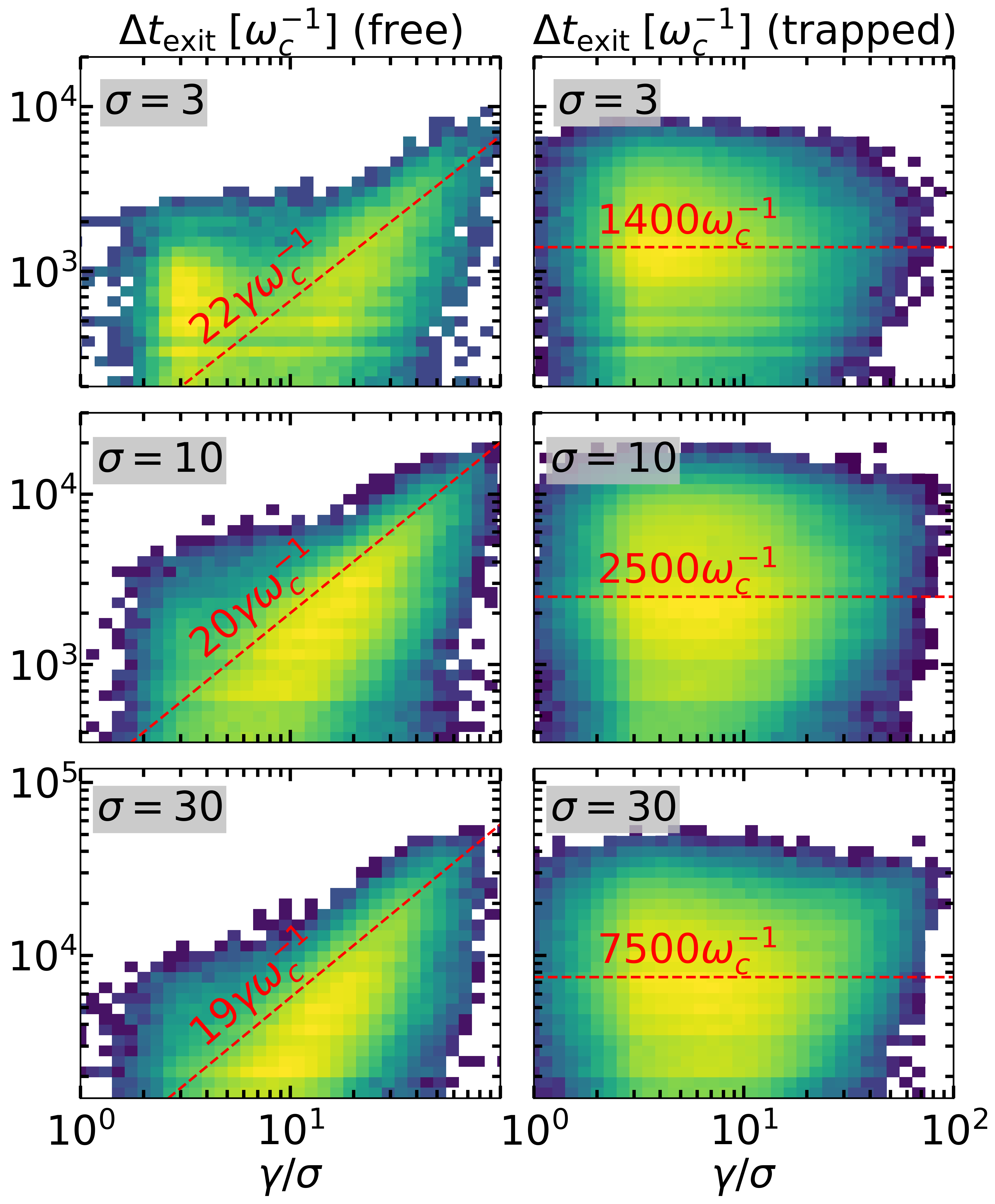}
    \caption{2D histograms of $\Delta t_{\rm exit}$. In the left column, we consider particles that exit the box as free, and define $\Delta t_{\rm exit}$ as the difference between the exit time $t_{\rm exit}$ and the time when their 
    last (i.e., current) free phase started.     
    In the right column, we  consider particles that exit the box as trapped, and define $\Delta t_{\rm exit}$ as the difference between the exit time $t_{\rm exit}$ and the time when their last free phase ended (i.e., the last time the particle got captured). Particles that never experienced a free phase are not included.  Dashed red lines on the left are the same as in the right panel of Fig.~5 in the main text.  Dashed red lines on the right indicate $L/c$. 
    The Lorentz factor $\gamma$ on the horizontal axis is measured at $t_{\rm exit}$. The histograms are normalized to their respective maxima, and colors span the range $[10^{-3},1]$ in logarithmic increments.
    }
    \label{fig:2dhistx}
\end{figure}

\subsection{The last free or trapped phase}
In the main text, we defined the escape time of free particles $\tesc$ as the duration of the free phase before they get trapped. Some of the free particles, though, will terminate their free phase of active acceleration not because they get trapped, but rather because they leave the simulation domain while still being free. The left column in \figg{2dhistx} shows, for particles that exit the box as free, the difference $\Delta t_{\rm exit}$ between their exit time $t_{\rm exit}$ and the time when the last (i.e., current) free phase started. Dashed red lines are the same as in the right column of Fig.~5 in the main text. The agreement of the dashed red lines with the locus of maxima of the 2D histograms suggests that the same estimate of $\tesc$ as in the main text can be used {\it both} for free particles that end up trapped (as in the main text) {\it and} for free particles that leave the system.

In the right column of \figg{2dhistx} we consider particles that exit the box as trapped, and define $\Delta t_{\rm exit}$ as the difference between their exit time $t_{\rm exit}$ and the time when the last free phase ended (i.e., the last time the particle got trapped). Regardless of $\gamma$, they spend $\sim L/c$ in their last trapped phase, before exiting the system. Particles that never experienced a free phase are not included.

\end{document}